\documentclass[showpacs,showkeys,preprint,twocolumn,amsmath,amssymb]{revtex4}
\usepackage{graphicx}
\usepackage{dcolumn}
\usepackage{bm}
\def\cE{\mathcal{E}}
\def\cF{\mathcal{F}}

\begin{document}
\title{On the corrections to the Casimir effect depending on the resolution of measurement}
\author{M.V.Altaisky} 
\affiliation{Joint Institute for Nuclear Research, Joliot Curie 6, Dubna, 141980, Russia; and Space Research Institute RAS, Profsoyuznaya 84/32, Moscow, 117997, Russia}
\email{altaisky@mx.iki.rssi.ru}
 
\author{N.E.Kaputkina} 
\affiliation{National  University of Science and Technology ``MISiS'', Leninsky prospect 4, Moscow, 119049, Russia} 
\email{nataly@misis.ru}

\date{Mar 30, 2011: Revised July 16, 2011}

\begin{abstract}
The Casimir force $\cF = -\frac{\pi^2\hbar c}{240a^4}$, which attracts to each other two perfectly conducting parallel plates separated by the distance $a$ in vacuum, is one of the blueprints of the reality of vacuum fluctuations. Following the recent conjecture, that quantum fields should 
be described in terms of the fields depending on the resolution of measurement, 
rather than the position  alone (M.V.Altaisky, {\em Phys. Rev. D} 81(2010)125003),
we derive the correction to the Casimir energy depending on the ratio of the plate displacement amplitude to the distance between plates. 
\end{abstract}
\maketitle
The Casimir force is the result of the difference of the vacuum zero point 
energy of the two different configurations: the rectangular volume  
$L_x\times L_y \times a$ bounded by two parallel 
conducting walls, and that of the same volume not bounded by conducting 
walls. In the former case, the electromagnetic field bounded between 
conducting walls is said to be dimensionally quantized, while in the 
latter case the frequency spectrum is continuous. The energy difference 
between these two configurations cannot be measured directly, but it varies 
with the variation of the gap $a$, and this variation can 
be measured as the Casimir force.

In 1948 Casimir conjected that the force between two parallel conducting 
planes depends only on two universal constants, $\hbar$ and $c$, and the 
distance between the plates $a$ \cite{Casimir1948}. The first attempt to 
measure the Casimir force has been undertaken in 1958 \cite{Sparnaay1958}.
Later the Casimir force have been measured with an atomic force microscope 
\cite{MR1998,BCOR2002}. A lot of studies related to the Casimir effect are being carried 
out in different branches of nanomechanics and photonics now, 
see e.g. \cite{BMM2007,RCG2011} and references therein for recent review.

In the dimensionally quantized case the zero point energy of the 
electromagnetic field between conducting plates is 
\begin{eqnarray} \nonumber 
E_Q &=& \frac{\hbar c}{2}\sum_\alpha |k_\alpha| =   \frac{\hbar c}{2}
\int  \frac{L_x L_y dk_\parallel^2}{(2\pi)^2}
\Bigl[|k_\parallel| +\\
&+& 2\sum_{n=1}^\infty \left(k_\parallel^2+\frac{\pi^2 n^2}{a^2} \right)^{1/2} 
\Bigr], \label{eq}
\end{eqnarray} 
where factor $2$ with the sum over discrete spectra accounts for two possible 
polarizations of the electromagnetic field; $k_\parallel \equiv (k_x,k_y)$. 
The energy of the same field free of any boundary conditions is expressed as 
 integral over continuous spectra 
\begin{equation}
E_0 =   \frac{\hbar c}{2}
\int \frac{ L_x L_y dk_\parallel^2}{(2\pi)^2} \int_{-\infty}^\infty
2\frac{dk_z}{(2\pi)}\left(k_\parallel^2+k_z^2\right)^{1/2}
\label{e0}.
\end{equation}
Both integrals (\ref{eq},\ref{e0}) are evidently infinite, but their difference
\begin{equation}
\cE = \frac{E_Q-E_0}{L_x L_y}, \label{ce}
\end{equation} 
known as the Casimir energy, can be regularized if the r.h.s. of the equations 
(\ref{eq},\ref{e0}) are multiplied by some cutoff function $f(k)$, such 
that $f(0)=1$ and $f\left(k\gg \frac{1}{a_0}\right) \to 0$, where $1/a_0$ 
is the inverse size of atom. This specific choice accounts for the fact, 
that the walls are metallic plates made of real atoms, rather than of an 
ideal conductor.

\begin{figure}[ht]
\centering \includegraphics[width=0.4\textwidth]{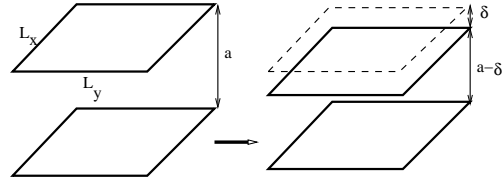}
\caption{Shift of the upper conducting wall from $z=a$ to $z=a-\delta$ 
changes the Casimir energy from $\cE(a,\delta)$ to $\cE(a-\delta,\delta)$ }
\label{m:pic}
\end{figure}

Such choice of the cutoff function $f$ is practically appropriate, but 
does not relate the calculated ultra-violet infinities to what is really 
measured at the finite scale of measurement $\delta$, see Fig.~\ref{m:pic}.
The conjecture, relating the quantum fields to the aperture function 
of the measurement taken with resolution $\delta$ was given in 
\cite{AltaiskyPRD10}. The 
aperture function $g(x)=-xe^{-x^2/2}$ leads 
in one-loop approximation, up to appropriate rescaling, to the cutoff function 
\begin{equation}
f(k)=e^{-4\delta^2 k^2}. \label{fk}
\end{equation}
After the choice \eqref{fk} the regularized Casimir energy is (see \S 3.2.4 of \cite{IZ1}):
\begin{eqnarray}\nonumber 
\cE &=&  \frac{\hbar c\pi^2}{4a^3}
\Bigl[
\frac{F(0)}{2}+F(1)+ F(2) +\ldots \\
\nonumber    &-& \int_0^\infty dn F(n) \Bigr],   \\  
\nonumber F(n) &=&   \int_0^\infty du \sqrt{u+n^2} 
e^{-4\pi^2\frac{\delta^2}{a^2}(u+n^2)} \\  
\label{cefn} &=&\frac{\sqrt\pi}{2 \left(\frac{2\pi\delta}{a} \right)^3}
\Bigl[
1 - {\rm erf} \left(\frac{2\pi\delta}{a} n  \right) +\\ 
\nonumber &+& 4\sqrt{\pi} \exp\left(-\left(\frac{2\pi\delta}{a}\right)^2n^2\right)\frac{\delta}{a} n 
\Bigr].
\end{eqnarray}
The difference between the sum and the integral in \eqref{cefn} is evaluated 
by Euler-Maclaurin formula
\begin{eqnarray*}
\frac{1}{2}F(0)+F(1)+ \ldots - \int_0^\infty dn F(n) = \\
=-\frac{1}{2!} B_2 F'(0) - \frac{1}{4!} B_4 F'''(0) 
- \ldots ,
\end{eqnarray*}
where $B_n$ are the Bernoulli numbers. 
This gives the corrections to the Casimir energy 
\begin{eqnarray}\nonumber 
\cE(a,\delta) &=& - \frac{\hbar c \pi^2}{720 a^3}
\Bigl[
1 + \frac{2}{7}\left(\frac{2\pi\delta}{a} \right)^2 + \\ 
&+& \frac{3}{28}\left(\frac{2\pi\delta}{a} \right)^4 + 
\ldots 
\Bigr],
\end{eqnarray}
and the Casimir force
\begin{eqnarray}\nonumber 
\cF(a,\delta) &=& - \frac{\hbar c \pi^2}{240 a^4}
\Bigl[
1 + \frac{10}{21}\left(\frac{2\pi\delta}{a} \right)^2 + \\
&+& \frac{1}{4}\left(\frac{2\pi\delta}{a} \right)^4 + 
\ldots 
\Bigr], \label{cfd}
\end{eqnarray}
respectively. 

We would like to emphasize, that if the conjecture of the previous paper \cite{AltaiskyPRD10}
is physically correct, and so the resolution of measurement $\delta$ is a 
real physical parameter, which constraints maximal momenta of the field 
fluctuations, rather than being a formal regularization parameter, 
the deviations from the standard results should be observed if we compare 
two measurements with the same gap between planes, but different resolution.

\begin{figure}[ht]
\centering \includegraphics[width=0.35\textwidth,angle=270]{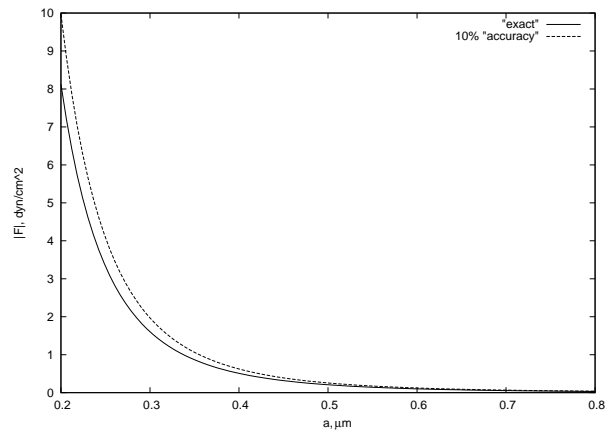}
\caption{Deviation of Casimir force between two plates of unit area in vacuum.
The solid line corresponds to the ``exact'' Casimir force ($\delta=0$), 
the dashed line corresponds to the scale-dependent Casimir force with $\delta/a=0.1$}
\label{ff:pic}
\end{figure}

In Fig.~\ref{ff:pic} below we present the comparison of the ``exact'' Casimir 
force between two plates in vacuum ($\delta=0$), and that calculated according to 
\eqref{cfd} with $\delta/a=0.1$.

The dependence of the Casimir force on the cutoff parameter have been already
suggested in the framework of the quantum field theory problem scaled to 
a condensed matter effective theory \cite{Volovik2001}, where the inter-atomic distance 
plays the same role as the Planck length in high energy physics. It was concluded that actual Casimir force should be stronger than 
that predicted by conventional Casimir theory. The dependence on the cutoff 
scale also raises some criticism against the application of regularization 
methods to Casimir effect, specially for spherical geometry \cite{Hagen2000}.

Interestingly, in present experiments the separation between plates is corrected by the 
factor $1+\left(\frac{\delta}{a}\right)^2$, derived from the Taylor expansion of the 
Casimir force, to account for the r.m.s. fluctuations of the random environment \cite{SKDL2011}. 
Our correction to the Casimir force due to the finite resolution of the 
measurement, given by Eq.\eqref{cfd} is also consistent with the limits posed 
by the precise measurement of the Casimir force given in \cite{Decca2007} 
with the resolution $\delta/a \approx 4\cdot 10^{-3}$.   
The choice of the scale parameter $\delta$ as a displacement amplitude is one of the possible 
simplifications. (Here we do not regard the dynamical effects \cite{FNL2005}.) For real experiments an important characteristics of the setup is a ratio of the boundary 
layer thickness to the distance between plates, which may be of order $h/a \sim 10^{-4} \div 10^{-2}$ 
\cite{MP2006}. With decreasing of the boundary plate thickness, according to the Lifshitz theory \cite{Lifshitz1956}, the electron plasma of metal boundaries become utmost transparent for high frequency photons \cite{Inui2004}, and 
the ratio $h/a$ plays a role of another cutoff parameter. 

An experimental study of the corrections to the Casimir force is certainly 
a challenging problem, where the dielectric permeability of the media 
should be taken into account 
at finite temperatures \cite{MT1997,BCOR2002,NLS2004}. 
\begin{acknowledgments}
The authors are thankful to prof. N.V.Antonov and  M.Z.Yuriev for useful discussions, and to prof. V.V.Nesterenko for critical comments. 
The research was supported in part by DFG Project 436 RUS 113/951 (M.V.A.) and by RFBR Project 11-02-00604-a and Programme of Creation and Development of the National University of Science and Technology "MISiS" (N.E.K.) 
\end{acknowledgments}

\begin{thebibliography}{18}
\expandafter\ifx\csname natexlab\endcsname\relax\def\natexlab#1{#1}\fi
\expandafter\ifx\csname bibnamefont\endcsname\relax
  \def\bibnamefont#1{#1}\fi
\expandafter\ifx\csname bibfnamefont\endcsname\relax
  \def\bibfnamefont#1{#1}\fi
\expandafter\ifx\csname citenamefont\endcsname\relax
  \def\citenamefont#1{#1}\fi
\expandafter\ifx\csname url\endcsname\relax
  \def\url#1{\texttt{#1}}\fi
\expandafter\ifx\csname urlprefix\endcsname\relax\def\urlprefix{URL }\fi
\providecommand{\bibinfo}[2]{#2}
\providecommand{\eprint}[2][]{\url{#2}}

\bibitem[{\citenamefont{Casimir}(1948)}]{Casimir1948}
\bibinfo{author}{\bibfnamefont{H.}~\bibnamefont{Casimir}},
  \bibinfo{journal}{Proc. Kon. Ned. Akad. Wet.} \textbf{\bibinfo{volume}{51}},
  \bibinfo{pages}{793} (\bibinfo{year}{1948}).

\bibitem[{\citenamefont{Sparnaay}(1958)}]{Sparnaay1958}
\bibinfo{author}{\bibfnamefont{M.}~\bibnamefont{Sparnaay}},
  \bibinfo{journal}{Physica} \textbf{\bibinfo{volume}{24}},
  \bibinfo{pages}{751} (\bibinfo{year}{1958}).

\bibitem[{\citenamefont{Mohideen and Roy}(1998)}]{MR1998}
\bibinfo{author}{\bibfnamefont{U.}~\bibnamefont{Mohideen}} \bibnamefont{and}
  \bibinfo{author}{\bibfnamefont{A.}~\bibnamefont{Roy}},
  \bibinfo{journal}{Physical Review Letters} \textbf{\bibinfo{volume}{81}},
  \bibinfo{pages}{4549} (\bibinfo{year}{1998}).

\bibitem[{\citenamefont{Bressi et~al.}(2002)\citenamefont{Bressi, Carugno,
  Onofrio, and Ruoso}}]{BCOR2002}
\bibinfo{author}{\bibfnamefont{G.}~\bibnamefont{Bressi}},
  \bibinfo{author}{\bibfnamefont{G.}~\bibnamefont{Carugno}},
  \bibinfo{author}{\bibfnamefont{R.}~\bibnamefont{Onofrio}}, \bibnamefont{and}
  \bibinfo{author}{\bibfnamefont{G.}~\bibnamefont{Ruoso}},
  \bibinfo{journal}{Phys. Rev. Lett.} \textbf{\bibinfo{volume}{88}},
  \bibinfo{pages}{041804} (\bibinfo{year}{2002}).

\bibitem[{\citenamefont{Bordag et~al.}(2007)\citenamefont{Bordag, Mohideen, and
  Mostepanenko}}]{BMM2007}
\bibinfo{author}{\bibfnamefont{M.}~\bibnamefont{Bordag}},
  \bibinfo{author}{\bibfnamefont{U.}~\bibnamefont{Mohideen}}, \bibnamefont{and}
  \bibinfo{author}{\bibfnamefont{V.}~\bibnamefont{Mostepanenko}},
  \bibinfo{journal}{Phys. Rep.} \textbf{\bibinfo{volume}{353}},
  \bibinfo{pages}{51} (\bibinfo{year}{2007}).

\bibitem[{\citenamefont{Rodriguez et~al.}(2011)\citenamefont{Rodriguez,
  Capasso, and Johnson}}]{RCG2011}
\bibinfo{author}{\bibfnamefont{A.}~\bibnamefont{Rodriguez}},
  \bibinfo{author}{\bibfnamefont{F.}~\bibnamefont{Capasso}}, \bibnamefont{and}
  \bibinfo{author}{\bibfnamefont{S.}~\bibnamefont{Johnson}},
  \bibinfo{journal}{Nature Photonics} \textbf{\bibinfo{volume}{5}},
  \bibinfo{pages}{211} (\bibinfo{year}{2011}).

\bibitem[{\citenamefont{Altaisky}(2010)}]{AltaiskyPRD10}
\bibinfo{author}{\bibfnamefont{M.}~\bibnamefont{Altaisky}},
  \bibinfo{journal}{Phys. Rev. D} \textbf{\bibinfo{volume}{81}},
  \bibinfo{pages}{125003} (\bibinfo{year}{2010}).

\bibitem[{\citenamefont{Itzykson and Zuber}(1980)}]{IZ1}
\bibinfo{author}{\bibfnamefont{C.}~\bibnamefont{Itzykson}} \bibnamefont{and}
  \bibinfo{author}{\bibfnamefont{J.-P.} \bibnamefont{Zuber}},
  \emph{\bibinfo{title}{Quantum field theory}}
  (\bibinfo{publisher}{McGraw-Hill, Inc.}, \bibinfo{year}{1980}).

\bibitem[{\citenamefont{Volovik}(2001)}]{Volovik2001}
\bibinfo{author}{\bibfnamefont{G.}~\bibnamefont{Volovik}},
  \bibinfo{journal}{JETP Lett.} \textbf{\bibinfo{volume}{73}},
  \bibinfo{pages}{375} (\bibinfo{year}{2001}).

\bibitem[{\citenamefont{Hagen}(2000)}]{Hagen2000}
\bibinfo{author}{\bibfnamefont{C.}~\bibnamefont{Hagen}},
  \bibinfo{journal}{Phys. Rev. D} \textbf{\bibinfo{volume}{61}},
  \bibinfo{pages}{065005} (\bibinfo{year}{2000}).

\bibitem[{\citenamefont{Sushkov et~al.}(2011)\citenamefont{Sushkov, Kim,
  Dalvit, and Lamoreaux}}]{SKDL2011}
\bibinfo{author}{\bibfnamefont{A.}~\bibnamefont{Sushkov}},
  \bibinfo{author}{\bibfnamefont{W.}~\bibnamefont{Kim}},
  \bibinfo{author}{\bibfnamefont{D.}~\bibnamefont{Dalvit}}, \bibnamefont{and}
  \bibinfo{author}{\bibfnamefont{S.}~\bibnamefont{Lamoreaux}},
  \bibinfo{journal}{Nature Physics} \textbf{\bibinfo{volume}{7}},
  \bibinfo{pages}{230} (\bibinfo{year}{2011}).

\bibitem[{\citenamefont{Decca et~al.}(2007)\citenamefont{Decca, L\'opez,
  Fischbach, Klimchitskaya, Krause, and Mostepanenko}}]{Decca2007}
\bibinfo{author}{\bibfnamefont{R.~S.} \bibnamefont{Decca}},
  \bibinfo{author}{\bibfnamefont{D.}~\bibnamefont{L\'opez}},
  \bibinfo{author}{\bibfnamefont{E.}~\bibnamefont{Fischbach}},
  \bibinfo{author}{\bibfnamefont{G.~L.} \bibnamefont{Klimchitskaya}},
  \bibinfo{author}{\bibfnamefont{D.~E.} \bibnamefont{Krause}},
  \bibnamefont{and} \bibinfo{author}{\bibfnamefont{V.~M.}
  \bibnamefont{Mostepanenko}}, \bibinfo{journal}{Phys. Rev. D}
  \textbf{\bibinfo{volume}{75}}, \bibinfo{pages}{077101}
  (\bibinfo{year}{2007}).

\bibitem[{\citenamefont{Fedotov et~al.}(2005)\citenamefont{Fedotov, Narozhny,
  and Lozovik}}]{FNL2005}
\bibinfo{author}{\bibfnamefont{A.}~\bibnamefont{Fedotov}},
  \bibinfo{author}{\bibfnamefont{N.}~\bibnamefont{Narozhny}}, \bibnamefont{and}
  \bibinfo{author}{\bibfnamefont{Y.}~\bibnamefont{Lozovik}},
  \bibinfo{journal}{Journal of Optics B: Quantum and Semiclassical Optics}
  \textbf{\bibinfo{volume}{7}}, \bibinfo{pages}{S64} (\bibinfo{year}{2005}).

\bibitem[{\citenamefont{Markov and Pis'mak}(2006)}]{MP2006}
\bibinfo{author}{\bibfnamefont{V.}~\bibnamefont{Markov}} \bibnamefont{and}
  \bibinfo{author}{\bibfnamefont{Y.~M.} \bibnamefont{Pis'mak}},
  \bibinfo{journal}{J. Phys. A: Math. Gen.} \textbf{\bibinfo{volume}{39}},
  \bibinfo{pages}{6525} (\bibinfo{year}{2006}).

\bibitem[{\citenamefont{Lifshitz}(1956)}]{Lifshitz1956}
\bibinfo{author}{\bibfnamefont{E.}~\bibnamefont{Lifshitz}},
  \bibinfo{journal}{Sov. Phys. JETP} \textbf{\bibinfo{volume}{2}},
  \bibinfo{pages}{73} (\bibinfo{year}{1956}).

\bibitem[{\citenamefont{Inui}(2004)}]{Inui2004}
\bibinfo{author}{\bibfnamefont{N.}~\bibnamefont{Inui}},
  \bibinfo{journal}{Journal of the Physical Society of Japan}
  \textbf{\bibinfo{volume}{73}}, \bibinfo{pages}{332} (\bibinfo{year}{2004}).

\bibitem[{\citenamefont{Mostepanenko and Trunov}(1997)}]{MT1997}
\bibinfo{author}{\bibfnamefont{V.}~\bibnamefont{Mostepanenko}}
  \bibnamefont{and} \bibinfo{author}{\bibfnamefont{N.}~\bibnamefont{Trunov}},
  \emph{\bibinfo{title}{The Casimir Effect and its Applications}}
  (\bibinfo{publisher}{Oxford University Press}, \bibinfo{address}{Oxford},
  \bibinfo{year}{1997}).

\bibitem[{\citenamefont{Nesterenko et~al.}(2004)\citenamefont{Nesterenko,
  Lambiase, and Scarpetta}}]{NLS2004}
\bibinfo{author}{\bibfnamefont{V.}~\bibnamefont{Nesterenko}},
  \bibinfo{author}{\bibfnamefont{G.}~\bibnamefont{Lambiase}}, \bibnamefont{and}
  \bibinfo{author}{\bibfnamefont{G.}~\bibnamefont{Scarpetta}},
  \bibinfo{journal}{Riv. Nuovo Cim.} \textbf{\bibinfo{volume}{27}},
  \bibinfo{pages}{1} (\bibinfo{year}{2004}).

\end{thebibliography}

\end{document}